%% file: ANA-EXOT-2022-33-PAPER.tex
\pdfoutput=1
\documentclass[PAPER, atlasdraft=false, UKenglish, cernpreprint, texmf, orcidlogo]{atlasdoc}
\usepackage{atlaspackage}
\usepackage{atlasbiblatex}

\usepackage{atlasphysics}

\usepackage{multirow}
\usepackage{siunitx}
\usepackage{overpic}
\usepackage{verbatim}
\usepackage{tabularx}
\newcolumntype{Y}{>{\centering\arraybackslash}X}
\usepackage{grffile} %
\usepackage[compat=1.0.0]{tikz-feynman}

\graphicspath{{logos/}{fig/public/}}

\usepackage{ANA-EXOT-2022-33-PAPER-defs}

\AtlasTitle{Search for boosted low-mass resonances decaying into hadrons produced in association with a photon in $pp$ collisions at $\sqrt{s}=13$~TeV with the ATLAS detector}

\AtlasAbstract{%
Many extensions of the Standard Model, including those with dark matter particles, propose new mediator particles that decay into hadrons.
This paper presents a search for such low mass narrow resonances decaying into hadrons using 140~fb$^{-1}$ of proton--proton collision data recorded with the ATLAS detector at a centre-of-mass energy of 13~TeV.
The resonances are searched for in the invariant mass spectrum of large-radius jets with two-pronged substructure that are recoiling against an energetic photon from initial state radiation, which is used as a trigger to circumvent limitations on the maximum data recording rate.
This technique enables the search for boosted hadronically decaying resonances in the mass range 20--100~GeV hitherto unprobed by the ATLAS Collaboration.
The observed data are found to agree with Standard Model predictions and 95\% confidence level upper limits are set on the coupling of a hypothetical new spin-1 $Z'$ resonance with Standard Model quarks as a function of the assumed $Z'$-boson mass in the range between 20 and 200~GeV.
}

\PreprintIdNumber{CERN-EP-2024-188}

\AtlasJournal{JHEP}
\AtlasJournalRef{\JHEP 01 (2025) 099}
\AtlasDOI{https://doi.org/10.1007/JHEP01(2025)099}

\hypersetup{pdftitle={ATLAS document},pdfauthor={The ATLAS Collaboration}}

\begin{document}

\maketitle

\section{Introduction}
\label{sec:intro}
Many extensions of the Standard Model (SM) predict new mediator particles that couple to SM particles.
A particular class of spin-1 mediators are referred to as \Zp.
These \Zp mediators appear in dark matter models~\cite{Abercrombie:2015wmb,Boveia:2016mrp} among many others~\cite{Randall:1999vf,Anchordoqui:2008di,Cullen:2000ef,Frampton:1987dn,Chivukula:2011ng,Simmons:1996fz,Baur:1989kv}.
The minimal requirement for a \Zp resonance to be produced in the $s$-channel at the LHC is a $\Zp\!\qqbar$ coupling, which in turn predicts resonances decaying into hadronic but not necessarily leptonic final states.
The first searches for hadronically decaying resonances at hadron colliders were carried out at the SPS by the UA1~\cite{UA1:1983zip,UA1:1988pgb} and UA2~\cite{UA2:1984tmm,UA2:1993tee} Collaborations.
The invariant mass reach was extended by the CDF~\cite{CDF:2008ieg} and D0~\cite{D0:2003lui} Collaborations at the Tevatron, and by the ATLAS~\cite{EXOT-2017-32,EXOT-2016-33,EXOT-2019-03} and CMS~\cite{CMS:2019gwf,CMS:2018kcg} Collaborations at the LHC. %
However, with the increase of the centre-of-mass energy and instantaneous luminosity, the sensitivity to light hadronically decaying resonances with small production cross sections is significantly hampered by the high QCD multijet production cross section,
which results in an event rate many orders of magnitude above the data recording bandwidth of the experiment.
This limitation has been partially addressed by performing searches using online reconstruction algorithms run at the trigger level~\cite{EXOT-2016-20,CMS:2018mgb}. Another technique is to use initial-state radiation (ISR)~\cite{An:2012ue} to avoid reliance on single-jet triggers with transverse momentum (\pt) thresholds of around $0.5~$TeV, which has been extensively used at the LHC~\cite{EXOT-2017-01,ATLAS:2024qqm,CMS:2019emo,CMS:2019xai,CMS:2019mcu}.

\begin{figure}[t]
\centering

\subfloat[]{\includegraphics[width=0.3\textwidth]{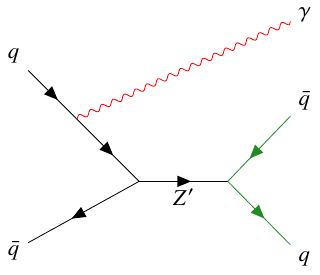}\label{fig:feyna}}
\hspace{0.5cm}
\subfloat[]{\includegraphics[width=0.3\textwidth]{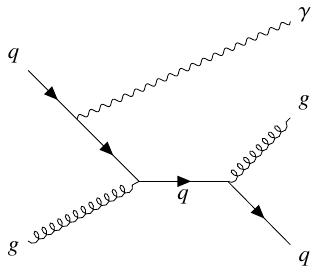}\label{fig:feynb}}
\hspace{0.5cm}
\subfloat[]{\includegraphics[width=0.3\textwidth]{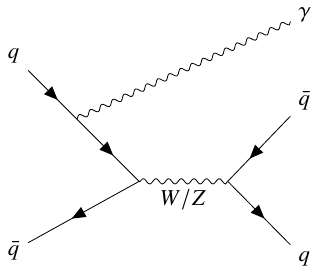}\label{fig:feync}}
\caption{
Representative leading-order Feynman diagrams for (a)~the $q\bar q \to\gamma+\Zp \to \gamma + q\bar q$ process, (b)~the non-resonant background from multijet production in association with a photon, and (c)~the dominant resonant background from \yv, where $V\to q\bar q$ and $V=W,Z$.
}
\label{fig:feyn}
\end{figure}

This paper presents a search for hadronically decaying resonances in the challenging invariant mass range between 20 and $100~\GeV$ that was hitherto unexplored by the ATLAS Collaboration. %
The search uses \lumi~fb$^{-1}$ of $pp$ collisions recorded by the ATLAS detector between 2015 and 2018 at a centre-of-mass energy of $\sqrt{s}=13~\TeV$.
The bandwidth and QCD background limitations are overcome by using a single-photon trigger with a transverse momentum threshold of $\pt > 140~\GeV$.
The search strategy selects ISR photons from the $q\bar q\to\Zp\to q\bar q$ process, resulting in a final state with hadronic activity from the $\Zp\to q\bar q$ decay recoiling against an energetic photon in the plane perpendicular to the beamline.
The leading-order Feynman diagram for the $q\bar q \to\gamma+\Zp \to \gamma + q\bar q$ process is shown in Figure~\ref{fig:feyna}.%

The Lorentz boost of the $\Zp$ boson brings an increased collimation of the $\Zp\to q\bar q$ decay products with decreasing \mZp, resulting in a very dense and experimentally challenging jet topology for $\mZp \lesssim 50~\GeV$.
This challenge is overcome by using the track-assisted reclustered (TAR) jet reconstruction technique~\cite{ATL-PHYS-PUB-2018-012} that combines information from the tracker and the calorimeters, and was first employed in Ref.~\cite{EXOT-2018-40}.
The decay products of the \Zp resonance are reconstructed as a single large-radius jet.%
The signal is searched for as a resonance in the spectrum of the invariant mass of the large-radius jet ($m_J$) over the background expected from SM processes.

The dominant backgrounds are non-resonant in \mj and arise from QCD multijet production in association with either a photon (see Figure~\ref{fig:feynb}) or a jet misidentified as a photon.
In both cases, QCD jets can pass the selection for a boosted massive $Z'$ resonance decaying into a pair of quarks due to rare fluctuations in parton showering and hadronisation folded with detector response.
Both background sources are estimated from data using signal-depleted control regions (CR) after validating the methodology with Monte Carlo (MC) simulations.
Resonant backgrounds arise from SM production of $\yv~(V=W,Z)$, with $V\to q\bar q$ (see Figure~\ref{fig:feync}).
Another resonant background arises from SM top-antitop-quark ($t\bar t$) production or associated production of a top-quark and a $W$ boson ($tW$), where one of the two $W$ bosons in the event decays into a quark pair and the other $W$ boson decays into an electron and a neutrino, with the former being misidentified as a photon.
All resonant backgrounds are estimated using MC simulations.
The correct modelling of the \mj distribution in resonant processes is verified in a dedicated CR enriched in \ttbar events.
The results are interpreted within the framework of a dark matter model with a \Zp mediator in the mass range between 20 and $200~\GeV$ that decays into a pair of quarks~\cite{Abercrombie:2015wmb,Boveia:2016mrp}.
The lower limit of the search range is chosen to avoid the non-trivial effects on the acceptance of the search, while the upper limit
is in the region where
other searches by the ATLAS Collaboration dominate the sensitivity~\cite{ATLAS:2024fdw,EXOT-2017-32,EXOT-2016-33,EXOT-2019-03,EXOT-2016-20,ATLAS:2024qqm,EXOT-2017-01}.

\section{ATLAS detector}
\label{sec:detector}

The ATLAS experiment~\cite{PERF-2007-01} at the LHC is a multipurpose particle detector
with a forward--backward symmetric cylindrical geometry and a near \(4\pi\) coverage in
solid angle.\footnote{\AtlasCoordFootnote}
It consists of an inner tracking detector surrounded by a thin superconducting solenoid
providing a \qty{2}{\tesla} axial magnetic field, electromagnetic and hadronic calorimeters, and a muon spectrometer.
The inner tracking detector covers the pseudorapidity range \(|\eta| < 2.5\).
It consists of silicon pixel, silicon microstrip, and transition radiation tracking detectors.
Lead/liquid-argon (LAr) sampling calorimeters provide electromagnetic (EM) energy measurements
with high granularity within the region \(|\eta|< 3.2\).
A steel/scintillator-tile hadronic calorimeter covers the central pseudorapidity range (\(|\eta| < 1.7\)).
The endcap and forward regions are instrumented with LAr calorimeters
for EM and hadronic energy measurements up to \(|\eta| = 4.9\).
The muon spectrometer surrounds the calorimeters and is based on
three large superconducting air-core toroidal magnets with eight coils each.
The field integral of the toroids ranges between \num{2.0} and \qty{6.0}{\tesla\metre}
across most of the detector.
The muon spectrometer includes a system of precision tracking chambers up to \(|\eta| = 2.7\) and fast detectors for triggering up to \(|\eta| = 2.4\).
The luminosity is measured mainly by the LUCID--2~\cite{LUCID2} detector, which is located close to the beampipe.
A two-level trigger system is used to select events~\cite{TRIG-2016-01}.
The first-level trigger is implemented in hardware and uses a subset of the detector information
to accept events at a rate below \qty{100}{\kHz}.
This is followed by a software-based trigger that
reduces the accepted event rate to \qty{1}{\kHz} on average
depending on the data-taking conditions.
A software suite~\cite{SOFT-2022-02} is used in data simulation, in the reconstruction
and analysis of real and simulated data, in detector operations, and in the trigger and data acquisition
systems of the experiment.

\section{Data and simulated event samples}
This analysis is performed using data from proton--proton ($pp$) collisions at $\sqrt{s}=13~\TeV$ at the LHC, collected during 2015--2018 with the ATLAS detector.
The total integrated luminosity of this data sample is $140~\ifb$ \cite{DAPR-2021-01},
obtained using the LUCID-2 detector for the primary luminosity measurements.
Data in this analysis are required to satisfy standard quality requirements~\cite{DAPR-2018-01}.

SM background processes and the \Zp signal are simulated using MC event generators. A detailed simulation of the ATLAS detector~\cite{SOFT-2010-01} based on the \GEANT~\cite{Agostinelli:2002hh} package is used to simulate the detector response for all MC event samples. Contributions from additional \pp interactions (pile-up) are simulated with the \PYTHIA~{8.186}~\cite{Sjostrand:2007gs} event generator using the \NNPDF[2.3lo] parton distribution function (PDF) set~\cite{Ball:2012cx} and corrected to match the spectrum of the average number of $pp$ collisions per bunch-crossing in the data.
Parton shower simulations with \PYTHIA use the A14 set of tuned parameters~\cite{ATL-PHYS-PUB-2014-021} with the \NNPDF[2.3lo] PDF set~\cite{Ball:2012cx} and EvtGen~\cite{Lange:2001uf} is used to model the decays of heavy-flavour hadrons.
All simulations using the \SHERPA event generator employ its internal parton shower model.

Prompt single-photon production was simulated with \SHERPA[2.2.1]~\cite{Bothmann:2019yzt}.
The parton-level process was generated at leading order (LO) in QCD for up to three additional partons, using the \CT[10nlo] PDF set~\cite{Lai:2010vv}, and matched to the parton shower using the \MEPSatLO prescription~\cite{Hoeche:2009rj}.
Photons from the matrix elements were required to be isolated according to a smooth-cone hadronic isolation criterion~\cite{Frixione:1998jh} with \(\delta_0=0.3\), \(\epsilon_{\gamma}=0.025\) and \(n=2\).
Multijet production was simulated using \PYTHIA[8.230]~\cite{Sjostrand:2014zea,Mrenna:2016sih} with the \NNPDF[2.3lo] PDF set with LO matrix elements for dijet production matched to the parton shower.
The renormalisation and factorisation scales were set to the geometric mean of the squared transverse masses of the two outgoing particles in the matrix element.
The \yv processes were simulated with \SHERPA[2.2.11]%
using the \NNPDF[3.0nnlo] PDF set~\cite{Ball:2014uwa}.
The perturbative calculations for \yv were performed at next-to-leading order (NLO) in QCD for up to one additional parton and LO for up to three additional partons, and matched to the parton shower~\cite{Gleisberg:2008fv,Schumann:2007mg}
using the MEPS@NLO prescription.
The matrix elements use the narrow-width approximation for the $V$ bosons.
Backgrounds from \ttbar and single-top-quark production were generated at \NLO in QCD with \POWHEGBOX[v2]~\cite{Frixione:2007nw,Nason:2004rx,Frixione:2007vw,Alioli:2010xd} using the \NNPDF[3.0nlo] PDF set~\cite{Ball:2014uwa} and interfaced to \PYTHIA[8.230] for parton showering and hadronisation.%
The diagram removal scheme~\cite{Frixione:2008yi} was used to remove interference and overlap between $tW$ and \ttbar production.
The \ttbar samples are normalized using calculations at NNLO in QCD including next-to-next-to-leading logarithmic corrections for soft-gluon radiation~\cite{Beneke:2011mq,Cacciari:2011hy,Baernreuther:2012ws,Czakon:2012zr,Czakon:2012pz,Czakon:2013goa,Czakon:2011xx}.
The single-top-quark processes are normalized to cross sections at \NLO in QCD from \textsc{Hathor}~v2.1~\cite{Aliev:2010zk,Kant:2014oha}.
The backgrounds from QCD \(V\)+jets production that are relevant for the top-quark control region defined in Section~\ref{sec:analysis} were simulated with \MGNLO[2.2.2]~\cite{Alwall:2014hca}, using LO-accurate matrix elements with up to four final-state partons.
The matrix-element calculation employed the \NNPDF[3.0nlo] set of PDFs, and was interfaced to \PYTHIA[8.186].
The overlap between matrix element and parton shower emissions was removed using the CKKW-L merging procedure~\cite{Lonnblad:2001iq,Lonnblad:2011xx}.
The \(V\)+jets samples were normalised to a next-to-next-to-leading-order
prediction in QCD~\cite{Anastasiou:2003ds}.

The signal model~\cite{Dobrescu:2013cmh,Backovic:2015soa,Abercrombie:2015wmb} features a vector resonance \Zp with the $\Zp\!\qqbar$  coupling $g_q$ set to 0.2 (with $q=u,d,s,c,b$), which results in a \Zp width well below the detector resolution.
The signal was simulated generating $\gamma+\Zp$ events at LO in QCD with \MGNLO[v2.9.2] using the \NNPDF[3.0nlo] PDF set and interfaced to \PYTHIA[8.244].
The rates for all decay modes except $\Zp\to\qqbar$ were set to 0, and the interference between the \Zp and the SM \Zboson boson was neglected~\cite{Abercrombie:2015wmb,Boveia:2016mrp}.
The translation of the results to other $g_q$ values is done following Ref.~\cite{Albert:2022xla}.
The mass of the hypothesised \Zp resonance ranged between $20\,\unit{\GeV}$ and $200\,\unit{\GeV}$. %
\section{Event reconstruction}
\label{sec:reconstruction}
At least one \pp collision vertex, reconstructed from at least two ID tracks, is required in the event~\cite{ATL-PHYS-PUB-2015-026}. The vertex whose associated tracks
give the highest sum of squared transverse momentum
is designated the primary vertex~(PV).
The ID tracks must have at least seven hits and satisfy $\pt>0.5~\GeV$ and $|\eta|<2.5$ requirements~\cite{ATL-PHYS-PUB-2015-006,ATL-PHYS-PUB-2015-051}. Their transverse and longitudinal impact parameters relative to the PV must satisfy $|\dzero|<2~\mathrm{mm}$ and $|\zzsth|<3~\mathrm{mm}$, respectively.

Photon candidates are reconstructed from three-dimensional topological clusters of energy deposits (topoclusters)
in the electromagnetic calorimeter~\cite{EGAM-2018-01}.
The contamination from neutral hadrons is reduced using the energy deposition profile in the first two calorimeter layers.
`Tight' criteria as defined in Ref.~\cite{PERF-2017-02} are applied for photon identification and isolation.
Photons are required to have \(\pt > \SI{10}{\GeV}\) and to fall within \(|\eta| < 1.37 \) or \(1.52 < |\eta| < 2.37\),
thus avoiding the transition region between the barrel and endcap calorimeters.

Jets are formed with the \antikt algorithm~\cite{Cacciari:2008gp,Fastjet}.
Jets with a radius parameter value of \(R=0.2\) are used in the overall reconstruction of the event.
These $R=0.2$ jets are built from topoclusters in the calorimeter that are corrected using the local cell signal weighting (LCW) method~\cite{PERF-2014-07}.
In addition, jets with \(R=0.4\) are used in the identification of $b$-hadron decays.
These $R=0.4$ jets are clustered from topoclusters and tracks with a particle flow algorithm that subtracts contributions due to charged particles from topoclusters~\cite{PERF-2015-09}.
A multivariate algorithm is used to identify \(R=0.4\) jets containing $b$-hadrons ($b$-tagging) with an average efficiency of 85\%~\cite{ATLAS:2022qxm}.%
Corrections for pile-up~\cite{Cacciari:2008gn} and to the energy scale and resolution~\cite{JETM-2018-05} are applied to both jet definitions.
Jets with \(R=0.2\) (\(R=0.4\)) are required to have \(\pt > \SI{25}{\GeV}\) and \(|\eta| < 2.2~(2.5)\). %
\(R=0.4\) jets with $\pt<60~\GeV$ and $|\eta|<2.4$ are identified as originating from the PV using associated tracks~\cite{PERF-2014-03}.
No such procedure is applied to $R=0.2$ jets since they are only used in association with tracks.

The TAR technique~\cite{ATL-PHYS-PUB-2018-012} is used to reconstruct the $\Zp \to q\bar q$ decay in the challenging boosted low-mass phase space for the first time.
This technique improves the resolution of jet substructure observables by combining tracking and calorimeter measurements.
The TAR jets are formed from ID tracks and \(R=0.2\) jets as follows.
The calibrated \(R=0.2\) jets are reclustered into larger jets with $R=1.0$
using trimming parameters optimized for ATLAS~\cite{JETM-2018-02}.
ID tracks are associated to the \(R=0.2\) subjets of the reclustered jet.
The \pt of each associated track is then rescaled by a common factor so the magnitude of the vector \pt sum of the associated tracks equals the \pt of the \(R=0.2\) jet.
These rescaled tracks are used to calculate TAR jet observables, e.g., the jet mass \mj.
Resonance candidate \(R=1.0\) TAR jets are selected to be central (\(|\eta| < 2\)),
have sufficient transverse momentum to balance the ISR photon (\(p_{\textrm{T},J} > \SI{150}{\GeV}\)),
and be collimated ($2\mj/p_{\textrm{T},J} < 1$). %
They must be separated from the highest-\pt photon in the event by \(\Delta \phi > \pi/2\),
and isolated from any photon with \(\pt^\gamma / \pt^J > 0.1\) by \(\Delta R \geq 1.2\).
TAR jets consisting of fewer than three rescaled tracks are rejected, since the substructure variable \(D_2\) introduced in Section~\ref{sec:analysis} is ill-defined for such jets~\cite{Larkoski:2013eya}.

Muons are reconstructed by matching a track or track segment found in the muon spectrometer to an ID track. Muons must satisfy the `Medium' requirements in Ref.~\cite{MUON-2018-03} and have $\pt>25~\GeV$ and $|\eta|<2.5$.
Muons must be isolated using the `Loose' criteria in Ref.~\cite{MUON-2018-03}, which limits the total energy observed in tracks and calorimeter deposits near the (extrapolated) muon track.

\section{Event selection and analysis strategy}
\label{sec:analysis}
The signal is characterised by hadronic activity from the $\Zp\to\qqbar$ decay produced back-to-back in azimuth to an isolated prompt energetic photon, which is used to trigger the event~\cite{TRIG-2018-05}.
To ensure the trigger is fully efficient, the transverse momentum of the leading photon \pty is required to be greater than \(\SI{150}{\GeV}\).
At least one TAR jet is required.
To reduce the non-resonant background contribution, events are rejected if any \(R=0.2\) jet within \(\Delta R \leq 1.2\)
of the leading TAR jet (but not reclustered into it) has a \pt greater than \pty.
Similarly, events where the leading TAR jet has $\rho\equiv\log(\mj^2/p_{\textrm{T},J}^2) \leq -5.4$ are rejected to improve the modelling of non-resonant backgrounds at the low end of the \mj spectrum using the data-driven approach, which will be described later in this Section.
This selection has a negligible effect on the signal acceptance in the targeted $\mZp$ range.

The internal energy distribution in TAR jets from signal events is consistent with a two-prong $\Zp\to\qqbar$ decay, while TAR jets from non-resonant backgrounds are compatible with a one-prong decay.
This difference is characterised using the \(D_2\) observable~\cite{Larkoski:2013eya}, which was found to be powerful at discriminating between one- and two-prong jets~\cite{Larkoski:2014gra}.
The $D_2$ variable is defined using ratios of energy correlation functions that explore the substructure of a jet using an angular-weighted sum over the momenta of its constituents.  The values of $D_2$ for two-prong jets are typically smaller than those for one-prong jets.
Figure~\ref{fig:D2-smoothed-cutmap:a} shows a comparison of the distribution of the $D_2$ observable between non-resonant background processes in three mass ranges and the resonant backgrounds.

The $D_2$ observable is correlated with \mj.
To eliminate the shaping of \mj resulting from the $D_2$ selection, a decorrelated observable~\cite{CMS-EXO-17-001,CMS-EXO-18-012,ATL-PHYS-PUB-2018-014,ATL-PHYS-PUB-2021-029,CMS-EXO-17-027} is defined as
\newcommand{\tagQuantile}{\ensuremath{D^{13\,\%}_{2}}\xspace}
\begin{equation}
\label{eq:DDT}
\twoprongtag(\rho, \pt) = D_2 - \tagQuantile(\rho, \pt)\,,
\end{equation}
where `DDT' stands for `Designed Decorrelated Tagger'.
Here, \tagQuantile is the 13\% quantile, i.e., the value of $D_2$ that splits the non-resonant background MC sample into two subsamples of 87\% and 13\%.
Hence, a selection of $\twoprongtag(\rho,\pt)<0$ has a fixed efficiency of 13\% in non-resonant background MC samples, irrespective of the mass or \pt of the jet.
The quantile value of 13\% was chosen by optimising the overall analysis sensitivity following the $S/\sqrt B$ metric, where $S$ and $B$ are the expected numbers of signal and background events, respectively, and by minimising the magnitude of the variations in the initial $\tagQuantile(\rho, \pt)$ distribution for non-resonant background events.

The final $\tagQuantile(\rho,\pt)$ map is obtained by smoothing its initial distribution by convolution with a two-dimensional Gaussian kernel~\cite{Altman:1992xxx}.
The parameters of the smoothing procedure like the binning of the initial $\tagQuantile(\rho,\pt)$ distribution and the kernel widths in $\rho$ and \pt dimensions are optimised by minimising the Jensen-Shannon distance~\cite{Lin:1991xxx,Kullback:1951xxx,Endres:2003xxx} between the \mj distributions for non-resonant background MC samples before and after the  $\twoprongtag(\rho,\pt)<0$ selection.
In addition to $D_2$, six other jet substructure observables capable of discriminating between one- and two-prong jets~\cite{Larkoski:2014gra}, including $\tau_{21}$ used in a similar previous ATLAS search~\cite{EXOT-2017-01}, were explored following the procedure above.
The $D_2$ variable was chosen since it minimises the shaping of the \mj distribution following the Jensen-Shannon distance metric.

Figure~\ref{fig:D2-smoothed-cutmap:b} shows the final $\tagQuantile(\rho,\pt)$ map after smoothing.
In the following, the `tagged' sample enriched in signal and two-prong resonant backgrounds is defined by the criterion \(\twoprongtag < 0\), while the complementary `anti-tagged' sample is enriched in non-resonant backgrounds.
\begin{figure}
\centering
\subfloat[]{\includegraphics[height=0.37\textwidth]{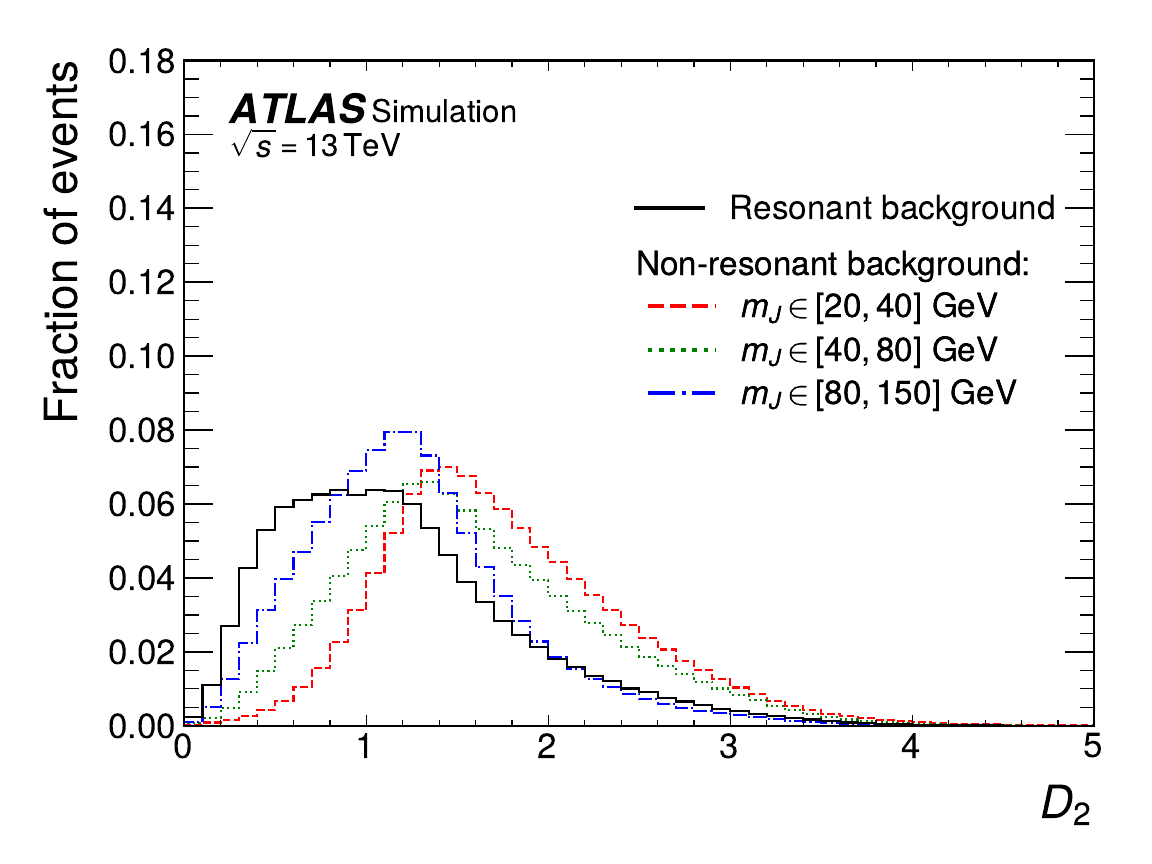}\label{fig:D2-smoothed-cutmap:a}}
\subfloat[]{\includegraphics[height=0.37\textwidth]{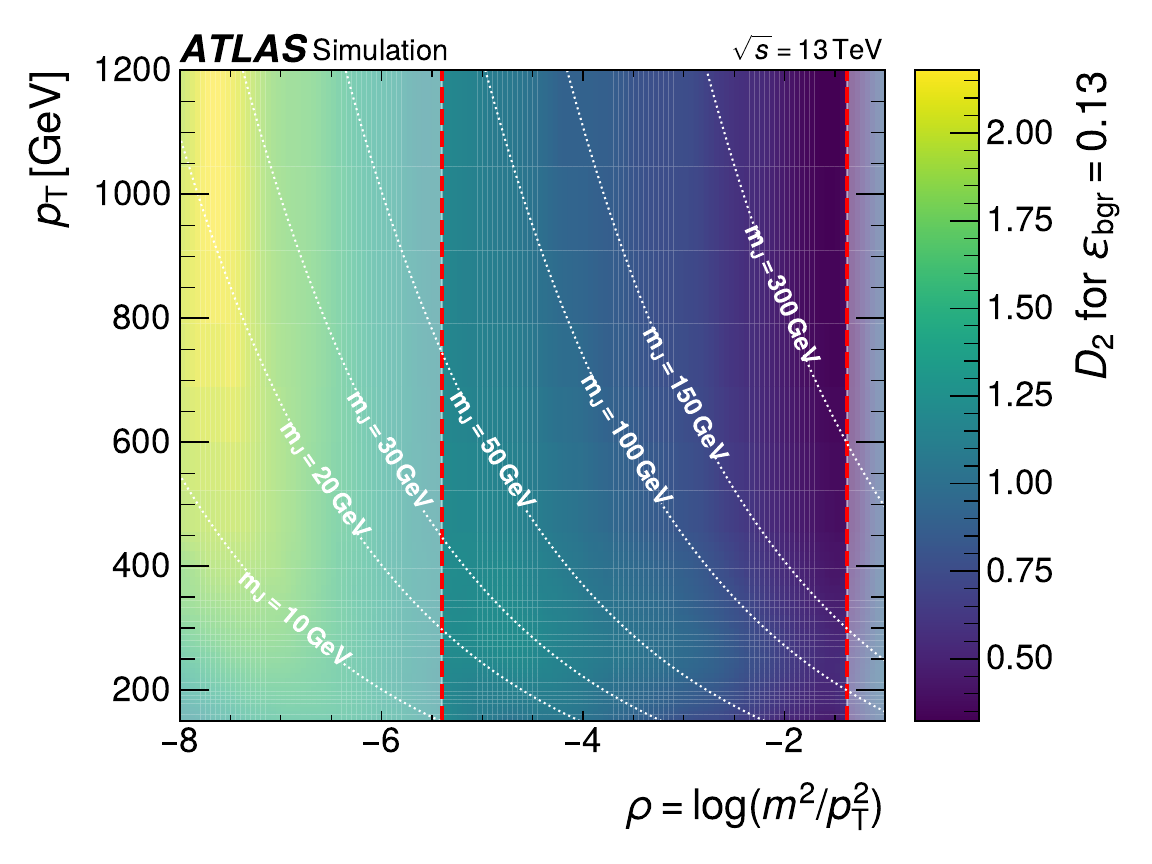}\label{fig:D2-smoothed-cutmap:b}}
\caption{
(a)~Distribution of the $D_2$ observable~\cite{Larkoski:2013eya} in MC simulations of non-resonant background processes in three different mass ranges (broken lines) compared to the resonant backgrounds (solid line).
(b)~The 13\% quantile of the $D_2$ observable in the simulated non-resonant background sample.
The contours of constant \mj are indicated by white dashed lines.
White-shaded areas separated by vertical broken lines indicate the phase space that the  \(\rho<-5.4\) and $2\mj/p_{\textrm{T},J} > 1$
selections reject.
}
\label{fig:D2-smoothed-cutmap}
\end{figure}

Ideally, the shape of the \mj distribution of non-resonant background should be identical in the tagged and anti-tagged samples.
In practice, the accuracy is limited by statistical and theoretical uncertainties on the modelling of non-resonant backgrounds in MC simulations.
Hence, this background distribution is estimated using data.
This is done by modifying the non-resonant background contribution taken from the anti-tagged region in data by an \mj-dependent correction $\ddtcorr(\mj)$ that is applied to the initial \ddt efficiency of 13\% as determined in simulations.
Typically, the $\ddtcorr(\mj)$ provides a percent-level correction to the nominal tagging efficiency.
In order to reduce the impact of statistical fluctuations, \ddtcorr is parameterized in \mj by a Bernstein polynomial~\cite{bib:bernstein} of order five.
In addition to providing an orthonormal basis, Bernstein polynomials bestow the advantage of strictly positive coefficients that are easily interpretable.
The order of the Bernstein polynomial was optimised using pseudoexperiments requiring that a spurious signal bias be smaller than one-third of the statistical uncertainty, and considering the $\chi^2$ probability, and the goodness-of-fit in data.
The six Bernstein coefficients of \ddtcorr are determined in the simultaneous fit to data described in Section~\ref{sec:fit}.

The data sample is further split into the central and forward regions defined by $|\eta_\gamma|\leq1.3$ and $|\eta_\gamma|>1.3$, respectively.
Together with the split into tagged and anti-tagged categories, this defines four analysis regions in total: `central tagged',  `forward tagged', `central anti-tagged', and `forward anti-tagged', where the former two are enriched in resonant processes due to the $\twoprongtag < 0$ selection.
The central tagged analysis region is then defined as the signal region (SR), since signal tends to be more central for such quark-induced processes with energetic ISR photons.
The other three analysis regions are referred to as control regions (CR).

\enlargethispage{5mm}

To minimize biases due to a potential signal contamination in the forward tagged CR and other CRs, a single $\ddtcorr(\mj)$ correction is determined and applied in all four regions in a simultaneous fit procedure discussed in Section~\ref{sec:fit}.
In all CRs, the signal contamination never exceeds three permille. Hence, the signal significance, defined as $S/\sqrt B$ in the \mj region around the \mZp candidate mass, is at least two times smaller in the CRs compared to the SR, at the level of sensitivity of this analysis.
A systematic uncertainty on the assumption that $\ddtcorr(\mj)$ is identical in the central and forward regions is estimated as described in Section~\ref{sec:fit}.
The resonant background processes (\yv, $tW$ and \ttbar) and the signal are modelled using MC simulations.
The efficiency of the \twoprongtag selection in MC simulations is calibrated using $W\to q\bar q'$ decays.
This is done using a dedicated top CR (denoted `top CR'), which targets the semi-muonic decay of \ttbar pairs and hence requires a single energetic muon trigger and exactly one isolated muon with \(\pt \geq \SI{30}{\GeV}\).
At least three \(R=0.4\) jets  must be present, of which at least two should be $b$-tagged. At least one $b$-tagged jet must fall within \(\Delta R < 1.5\) from the muon to select for collimated $t\to bW$ decays.
Furthermore, at least one TAR jet passing all SR selections except for \twoprongtag must be present.
TAR jets must be separated by \(\Delta R \geq 1.05\) from muons and by $\Delta R \geq 1.45$ from $b$-tagged jets.

The fiducial on-shell $\Wboson\rightarrow q q'$ production rate is extracted  in the top CR using a parametrised fit to the $W$-boson candidate mass spectrum, where the contribution from \Wboson bosons is captured by a Gaussian distribution, while the background is described by a second order polynomial.
The efficiency of the \twoprongtag selection $\varepsilon_{\twoprongtag}^{\text{res}}$ for resonant \Wboson-boson production is then determined as the ratio of the fiducial on-shell \Wboson-boson production rates after and before applying the \(\twoprongtag < 0\) requirement.
The ratio
\begin{equation*}
\mathcal{R}_{\twoprongtag}=\frac{ \varepsilon_{\twoprongtag,\text{data}}^{\text{res}} }{ \varepsilon_{\twoprongtag,\text{MC}}^{\text{res}} }
\end{equation*}
is then used to calibrate the tagging efficiency in MC simulations of resonant processes.
The calibration factor is measured to be $\mathcal{R}_{\twoprongtag} = 0.971\pm 0.026$.
Its value and the corresponding uncertainty is propagated to the signal extraction fit described in Section~\ref{sec:fit}.

\section{Statistical analysis and systematic uncertainties}
\label{sec:fit}
The resonant $\Zp \to q\bar q$ signal is extracted via a simultaneous maximum-likelihood fit~\cite{Moneta:2010pm,Verkerke:2003ir_}  to the binned $\mj$ distributions in the SR and the three CR categories, but not the top CR.
This fit considers signal and resonant background predictions from MC simulations scaled to their theoretical cross sections and calibrated for the \twoprongtag selection efficiency determined in the top CR.
The width of the bins in \mj is chosen as about one-third of the experimental resolution on \mZp while keeping the statistical uncertainties per bin approximately constant.
The expected yields $\nu_i$ for the Poisson probability density in a given \mj bin $i$ are given as
\begin{equation}
\label{eq:pd_SR}
\nu_{\text{SR},i}(\mu,\theta)
= \mu\cdot\nu_{\text{SR,sig},i}(\theta)
+ \sum_\text{res. bkg.}\nu_{\text{SR,bkg},i}(\theta)
+ \frac{13\%}{1-13\%}\ddtcorr{}_{,i}\cdot\nu_{\text{central},i}
\end{equation}
for the SR and
\begin{equation}
\label{eq:pd_cent_antitag_CR}
\nu_{\text{caCR},i}(\mu,\theta)
= \mu\cdot\nu_{\text{caCR,sig},i}(\theta)
+ \sum_\text{res. bkg}\nu_{\text{caCR,bkg},i}(\theta)
+ \nu_{\text{central},i}
\end{equation}
in the central anti-tagged CR labelled as `caCR'.
Here, $\mu$ is the signal strength, i.e. a factor multiplying the expected signal yields, $\theta$ is the vector of nuisance parameters representing systematic uncertainties and the $\mathcal{R}_{\twoprongtag}$ calibration, `sig' labels the signal and `res.~bkg' the resonant background processes.
The values $\nu_{\text{central},i}$ approximate the yield of the non-resonant background in the central anti-tagged CR.
The relevant difference between Eqs.~\eqref{eq:pd_SR} and \eqref{eq:pd_cent_antitag_CR} is the transfer factor $\frac{13\%}{1-13\%}\ddtcorr{}_{,i}$ that considers the non-resonant background tagging efficiency in data, as a function of \mj.
The probability densities in the forward tagged and anti-tagged CRs are defined in analogy to Eqs.~\eqref{eq:pd_SR} and \eqref{eq:pd_cent_antitag_CR}, respectively, using the same $\frac{13\%}{1-13\%}\ddtcorr{}_{,i}$ transfer factor.
Hence, the non-resonant background contribution is simultaneously fit in the SR and the three CRs through a common set of the Bernstein coefficients for the $\ddtcorr(\mj)$ correction.
The uncertainty in the non-resonant background estimate represents the dominant systematic uncertainty in this analysis, contributing between 80\% and 90\% of the total systematic uncertainty, depending on the \mZp hypothesis.
This uncertainty estimate considers three different sources, which are described below.
\\
First, a systematic uncertainty on the assumption that $\ddtcorr(\mj)$ is identical in the central and forward regions is explicitly considered in the simultaneous maximum-likelihood fit procedure through Eqs.~\eqref{eq:pd_SR} and \eqref{eq:pd_cent_antitag_CR} and their counterparts for the forward CRs.
Since the central and forward regions use the same $\ddtcorr{}$ correction factor and hence the same Bernstein coefficients, any tensions between the data in the tagged central and forward regions and the estimates obtained from the corresponding untagged regions will result in an increased uncertainty in the fitting procedure.
This source of systematic uncertainty explicitly accounts for statistical limitations due to a finite number of data and MC events, and contributes more than 90\% of the uncertainty on the non-resonant background estimate.
\\
Second, a systematic uncertainty accounting for a potential spurious signal bias as a function of \mj is considered.
This bias is evaluated through a multiplicative nuisance parameter on the signal yield in fits to pseudoexperiments derived from Asimov data.
The corresponding uncertainty is below one percent of the uncertainty on the non-resonant background estimate.
\\
Third, a systematic uncertainty related to the extrapolation between the forward and central regions is evaluated. This uncertainty accounts for localised fluctuations in tagged region data on scales that are comparable to the $\mj$ resolution.
This uncertainty is derived from the net difference between observed data and the pre-fit background prediction.
If this net difference is larger than $\sqrt B$ in a given \mj bin, a systematic uncertainty is added to this bin.
Only the tagged forward control region is used to determine the net difference in order to minimise any impact from signal that may potentially be present.
This systematic uncertainty is applied in the central and forward regions, per \mj bin and it is taken as uncorrelated across bins.
It contributes a few percent of the uncertainty in the non-resonant background estimate.

The systematic uncertainties related to MC simulations are parameterized as nuisance parameters with Gaussian or log-normal prior probabilities, are profiled and used to constrain the template shapes and the normalisations varied in the fit.
The leading sources of systematic uncertainty related to MC simulations originate from the theoretical modelling of signal events. %
Among those uncertainties are:
the choice of the factorisation and renormalisation scales,
the choice of PDFs,
and the uncertainty on the strong coupling constant $\alphas$.
Sub-leading sources of systematic uncertainties that affect both signal and background events modelled using MC simulations are related to the energy scale and resolution of $R=0.2$ and $R=0.4$ jets~\cite{JETM-2018-05}.
Other, typically negligible sources of systematic uncertainty are related to the identification and reconstruction of photons~\cite{EGAM-2018-01}, the finite number of MC events, and the measured integrated luminosity~\cite{DAPR-2021-01}.
Overall, this search is limited by statistical uncertainties in the data, which typically range between 20\% and 50\% of the theoretical cross section for the $\Zp \to q\bar q$ process, depending on $m_{\Zp}$.

\section{Results}
\label{sec:result}

The distributions of \mj in the SR and the central anti-tagged CR and are shown in Figure~\ref{fig:ms-body} after the fit to data under the background-only hypothesis.
The statistical weight of the anti-tagged CRs is significantly higher than that of the tagged regions since they have seven times the yield.
Hence, the excursion of the data points from the background prediction
relative to the total statistical plus systematic uncertainty
in the anti-tagged CRs is much smaller than in the tagged regions.
For the same reason, the relative contribution of the systematic uncertainty, shown as a hatched band, appears larger in the anti-tagged CRs due to the smaller total uncertainty compared to the tagged regions.
Figure~\ref{fig:tf} shows the corresponding $\ddtcorr(\mj)$ correction after the fit to data under the background-only hypothesis alongside observed ratios of tagged over anti-tagged events in the central and forward regions.
No significant deviation from SM predictions is observed.

\begin{figure}
\centering
\subfloat[]{\includegraphics[width=0.425\textwidth]{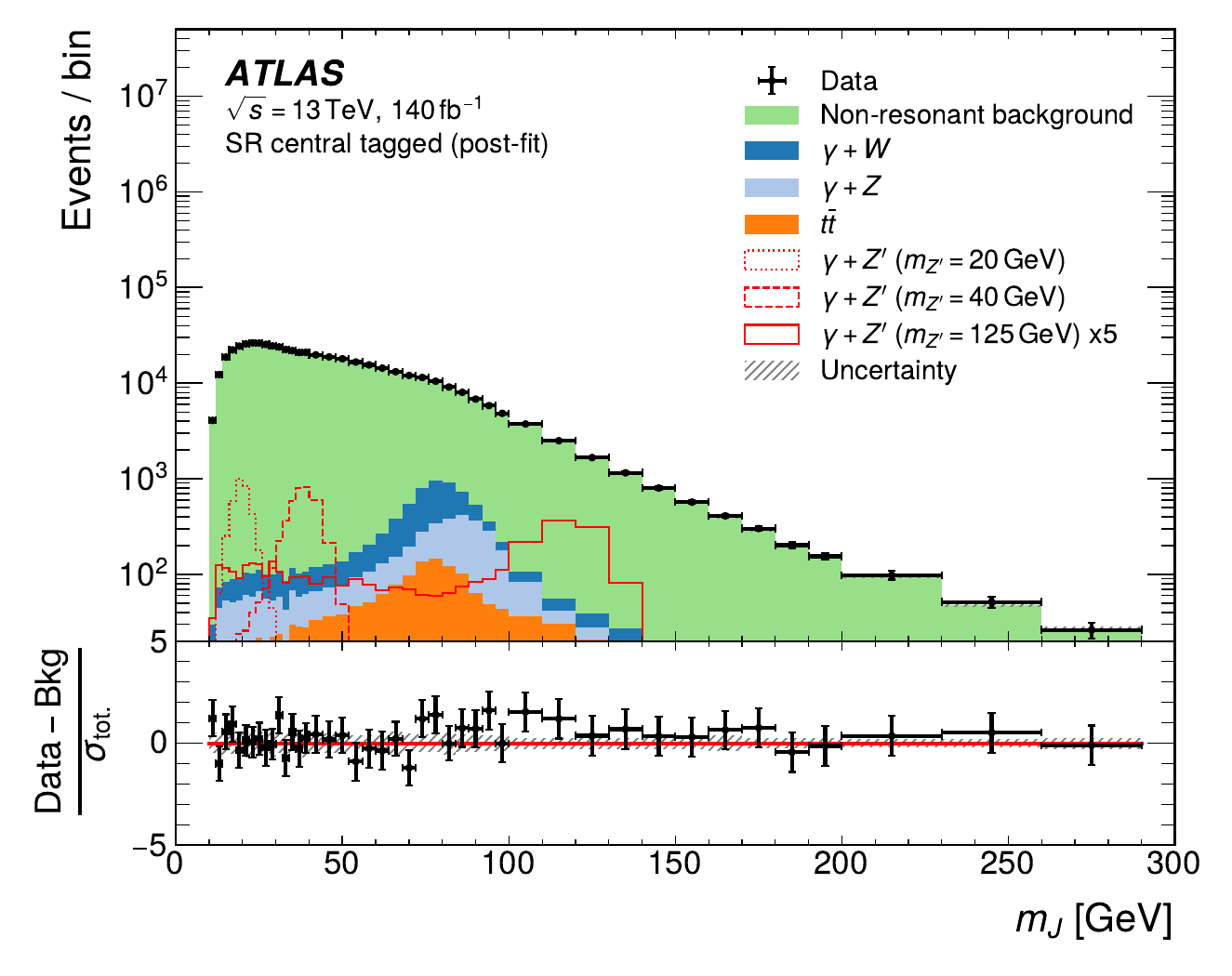}}
\subfloat[]{\includegraphics[width=0.425\textwidth]{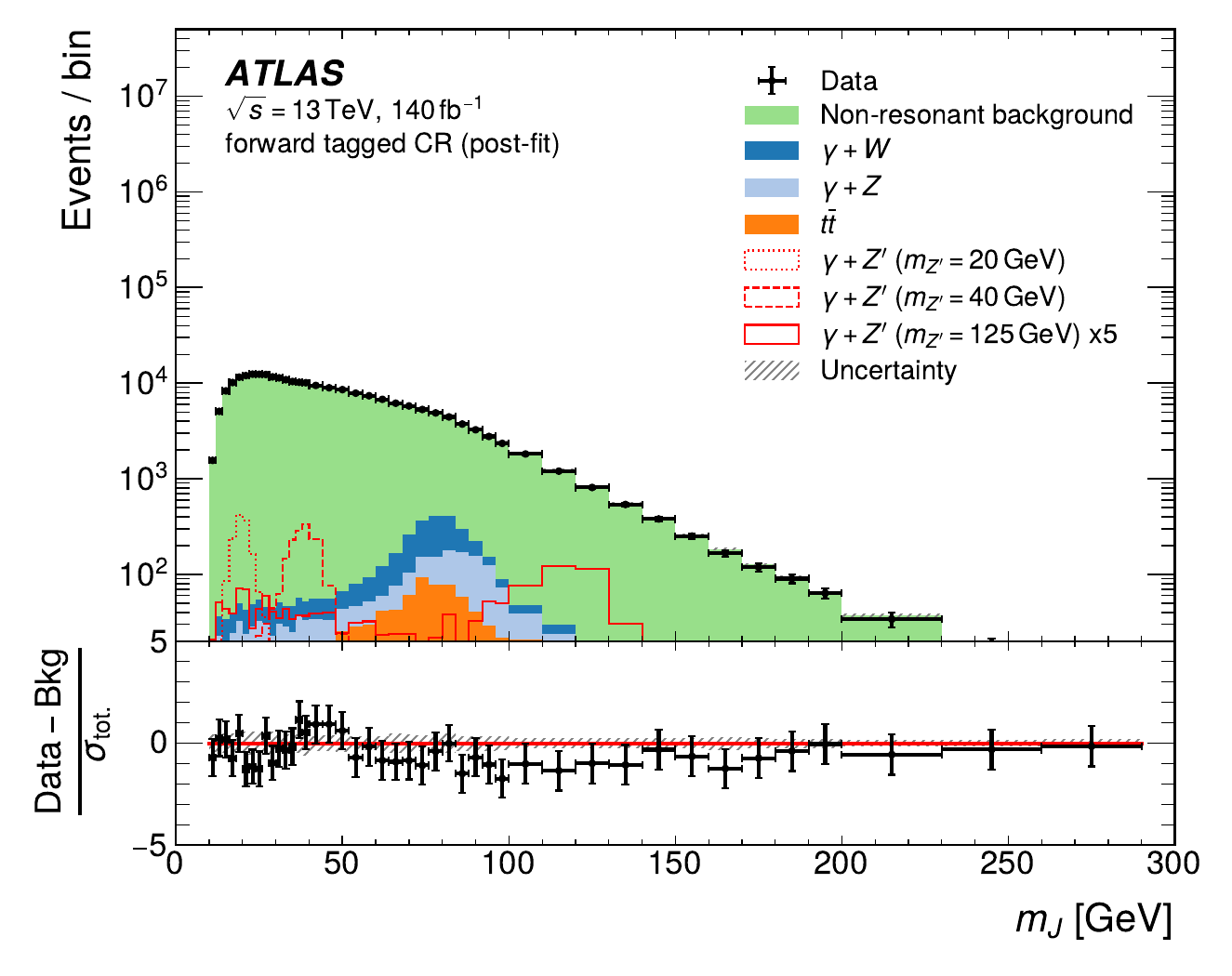}} \\
\subfloat[]{\includegraphics[width=0.425\textwidth]{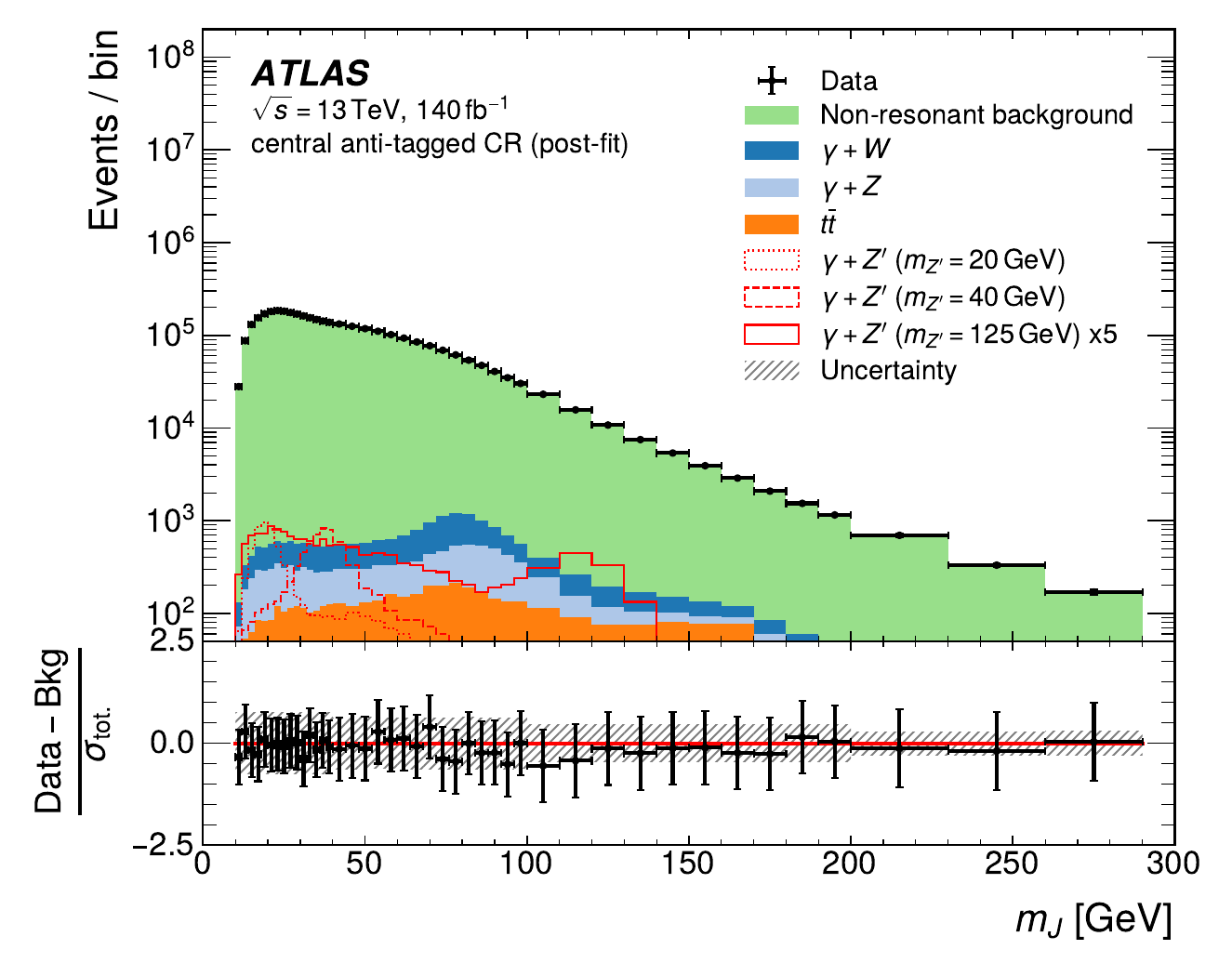}}
\subfloat[]{\includegraphics[width=0.425\textwidth]{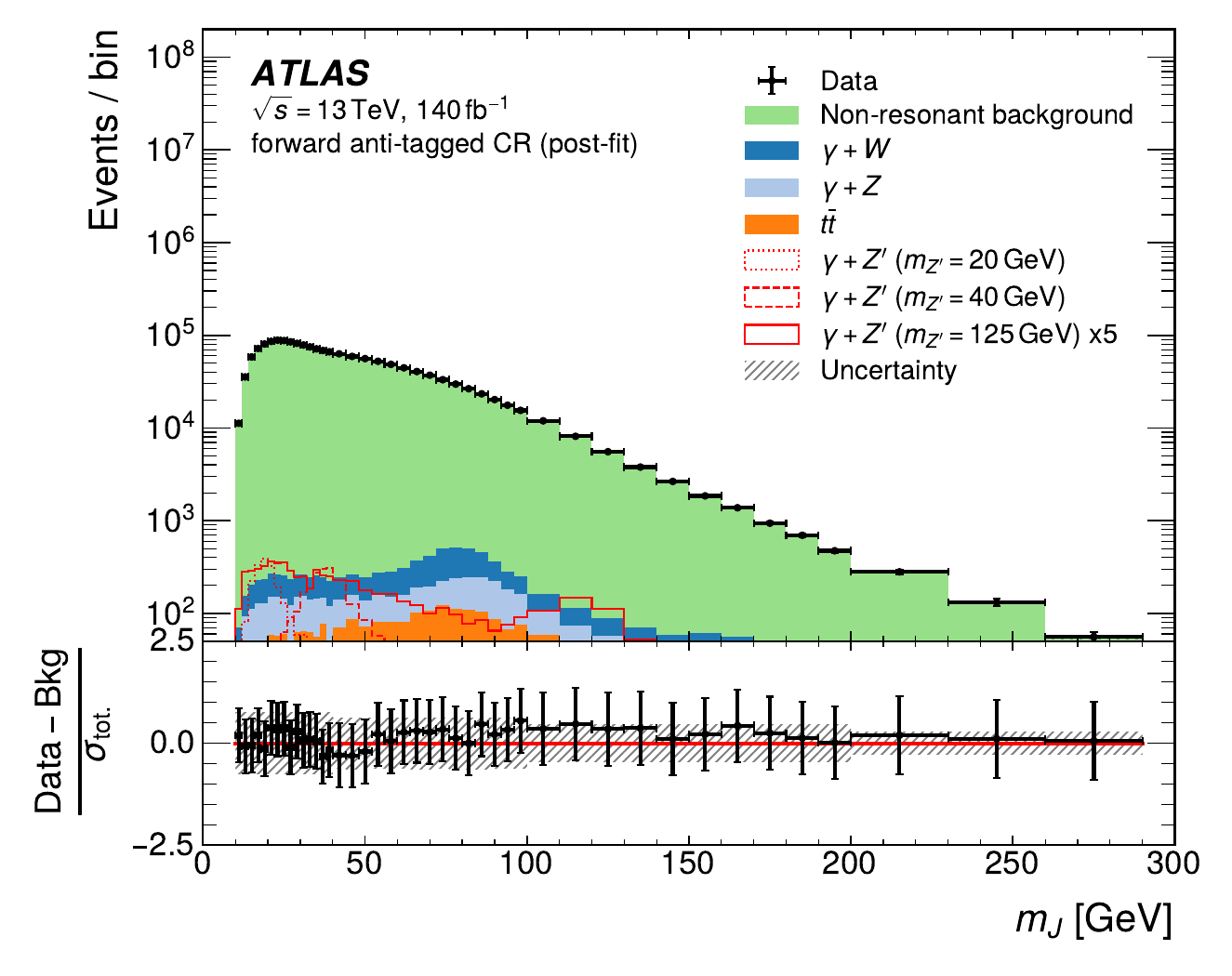}}
\caption{
Invariant mass \mj of the resonance candidates in the
(a)~SR, i.e., central tagged region,
(b)~forward tagged CR,
(c)~central anti-tagged CR, and
(d)~forward anti-tagged CR,
after the fit to data under the background-only hypothesis.
The total systematic uncertainty is shown as the hatched band.
Three representative \yz signal distributions are overlaid as red lines.
The signal is shown for $g_q=0.2$ with production cross sections of \(\SI{309}{fb}\), \(\SI{143}{fb}\),
and \(\SI{34.2}{fb}\)
for  \(m_\Zp=20,~50,~\text{and}~125~\GeV\), respectively.
}
\label{fig:ms-body}
\end{figure}

In absence of a significant excess, the results are interpreted in Figure~\ref{fig:limit_gq} as exclusion limits at 95\% confidence level (CL) on the $g_q$ coupling strength with the $CL_s$ formalism~\cite{Read:2002hq} using a profile likelihood ratio~\cite{Cowan:2010js} as test statistic.
In the range $\mZp<100~\GeV$, hitherto unprobed by ATLAS, $g_q$ couplings as small as 0.1 are excluded.

\begin{figure}[t]
\centering
\includegraphics[width=0.44\textwidth]{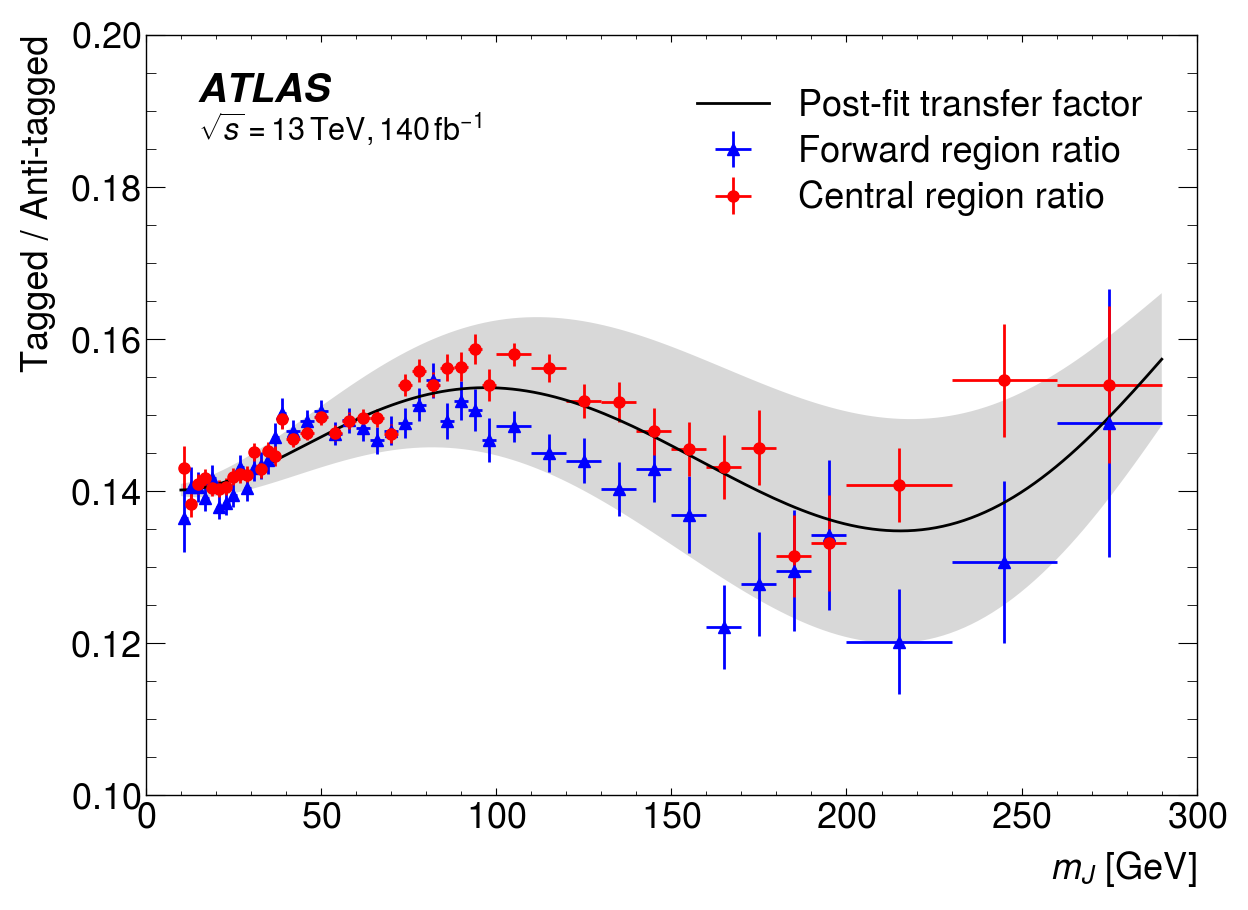}
\caption{
The distribution of the transfer factor $\frac{13\%}{1-13\%}\ddtcorr(\mj)$ after the fit to data under the background-only hypothesis.
The corresponding uncertainty band is shown as a shaded area.
Also shown are the observed ratios of tagged over anti-tagged events in the central~(red circles) and forward~(blue triangles) regions, where the prior expected contributions from resonant backgrounds have been subtracted and the uncertainties are purely statistical.}
\label{fig:tf}
\end{figure}

\begin{figure}[t]
\centering
\includegraphics[width=0.49\textwidth]{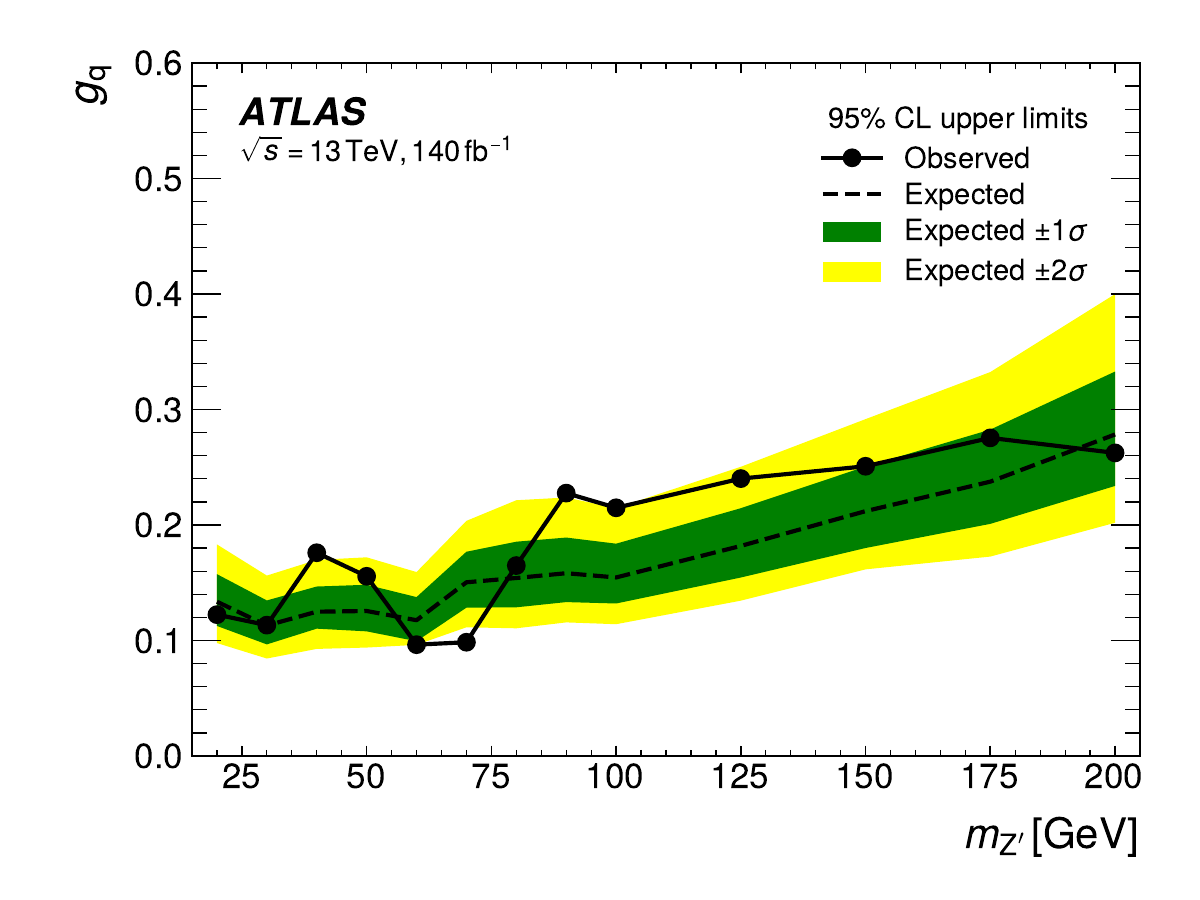}
\caption{
Observed (expected) upper exclusion limits at 95$\%$ CL on the coupling strength $g_q$ of a new vector \Zp particle decaying to a \qqbar pair for the LHC DM WG benchmark signal model from Refs.~\cite{Dobrescu:2013cmh,Backovic:2015soa,Abercrombie:2015wmb} where decay rates for all decay modes except $\Zp\to\qqbar$ are set to 0, and the interference between the \Zp and the SM \Zboson boson is neglected.
The expected $\pm1\sigma$ ($\pm2\sigma$) interval around the expected limit is shown as the filled green (yellow) band.
}
\label{fig:limit_gq}
\end{figure}

\FloatBarrier

\section{Conclusion}
\label{sec:conclusion}

This paper presents a search for light hadronically decaying resonances
using \lumi~fb$^{-1}$ of $pp$ collision data at $\sqrt{s}=13~\TeV$ recorded by the ATLAS detector at the LHC.
To avoid bandwidth limitations, events are triggered using energetic photons radiated from the initial state,
and a search for resonances is performed on the invariant mass of the recoiling hadronic system.
A track-assisted reclustered jet reconstruction method that combines information from the tracker and the calorimeters is used to reconstruct these boosted hadronically decaying resonances.
A data-driven method is used to estimate the non-resonant background from the Standard Model production of a photon in association with jets and QCD multijet production with a jet misidentified as a photon.
These techniques allow a search
for new hadronically decaying resonances with invariant masses from 20 to 100~\GeV for the first time in ATLAS.
No evidence for new resonances is found.  The results are interpreted within the framework of a dark matter model with a \Zp mediator, i.e., the $\qqbar\to \gamma + \Zp \to \gamma + \qqbar$ process, and probe previously uncharted
parameter space for
the $\Zp\qqbar$ coupling $g_q$ for masses
$20 < m_{\Zp} < 200~\GeV$,
excluding $g_q$ couplings down to 0.1.

\section*{Acknowledgements}
%

%
%

%
%

We thank CERN for the very successful operation of the LHC and its injectors, as well as the support staff at
CERN and at our institutions worldwide without whom ATLAS could not be operated efficiently.

The crucial computing support from all WLCG partners is acknowledged gratefully, in particular from CERN, the ATLAS Tier-1 facilities at TRIUMF/SFU (Canada), NDGF (Denmark, Norway, Sweden), CC-IN2P3 (France), KIT/GridKA (Germany), INFN-CNAF (Italy), NL-T1 (Netherlands), PIC (Spain), RAL (UK) and BNL (USA), the Tier-2 facilities worldwide and large non-WLCG resource providers. Major contributors of computing resources are listed in Ref.~\cite{ATL-SOFT-PUB-2023-001}.

We gratefully acknowledge the support of ANPCyT, Argentina; YerPhI, Armenia; ARC, Australia; BMWFW and FWF, Austria; ANAS, Azerbaijan; CNPq and FAPESP, Brazil; NSERC, NRC and CFI, Canada; CERN; ANID, Chile; CAS, MOST and NSFC, China; Minciencias, Colombia; MEYS CR, Czech Republic; DNRF and DNSRC, Denmark; IN2P3-CNRS and CEA-DRF/IRFU, France; SRNSFG, Georgia; BMBF, HGF and MPG, Germany; GSRI, Greece; RGC and Hong Kong SAR, China; ISF and Benoziyo Center, Israel; INFN, Italy; MEXT and JSPS, Japan; CNRST, Morocco; NWO, Netherlands; RCN, Norway; MNiSW, Poland; FCT, Portugal; MNE/IFA, Romania; MESTD, Serbia; MSSR, Slovakia; ARIS and MVZI, Slovenia; DSI/NRF, South Africa; MICIU/AEI, Spain; SRC and Wallenberg Foundation, Sweden; SERI, SNSF and Cantons of Bern and Geneva, Switzerland; NSTC, Taipei; TENMAK, T\"urkiye; STFC/UKRI, United Kingdom; DOE and NSF, United States of America.

Individual groups and members have received support from BCKDF, CANARIE, CRC and DRAC, Canada; CERN-CZ, FORTE and PRIMUS, Czech Republic; COST, ERC, ERDF, Horizon 2020, ICSC-NextGenerationEU and Marie Sk{\l}odowska-Curie Actions, European Union; Investissements d'Avenir Labex, Investissements d'Avenir Idex and ANR, France; DFG and AvH Foundation, Germany; Herakleitos, Thales and Aristeia programmes co-financed by EU-ESF and the Greek NSRF, Greece; BSF-NSF and MINERVA, Israel; NCN and NAWA, Poland; La Caixa Banking Foundation, CERCA Programme Generalitat de Catalunya and PROMETEO and GenT Programmes Generalitat Valenciana, Spain; G\"{o}ran Gustafssons Stiftelse, Sweden; The Royal Society and Leverhulme Trust, United Kingdom.

In addition, individual members wish to acknowledge support from Armenia: Yerevan Physics Institute (FAPERJ); CERN: European Organization for Nuclear Research (CERN PJAS); Chile: Agencia Nacional de Investigaci\'on y Desarrollo (FONDECYT 1230812, FONDECYT 1230987, FONDECYT 1240864); China: Chinese Ministry of Science and Technology (MOST-2023YFA1605700), National Natural Science Foundation of China (NSFC - 12175119, NSFC 12275265, NSFC-12075060); Czech Republic: Czech Science Foundation (GACR - 24-11373S), Ministry of Education Youth and Sports (FORTE CZ.02.01.01/00/22\_008/0004632), PRIMUS Research Programme (PRIMUS/21/SCI/017); EU: H2020 European Research Council (ERC - 101002463); European Union: European Research Council (ERC - 948254, ERC 101089007), Horizon 2020 Framework Programme (MUCCA - CHIST-ERA-19-XAI-00), European Union, Future Artificial Intelligence Research (FAIR-NextGenerationEU PE00000013), Italian Center for High Performance Computing, Big Data and Quantum Computing (ICSC, NextGenerationEU); France: Agence Nationale de la Recherche (ANR-20-CE31-0013, ANR-21-CE31-0013, ANR-21-CE31-0022, ANR-22-EDIR-0002), Investissements d'Avenir Labex (ANR-11-LABX-0012); Germany: Baden-Württemberg Stiftung (BW Stiftung-Postdoc Eliteprogramme), Deutsche Forschungsgemeinschaft (DFG - 469666862, DFG - CR 312/5-2); Italy: Istituto Nazionale di Fisica Nucleare (ICSC, NextGenerationEU), Ministero dell'Università e della Ricerca (PRIN - 20223N7F8K - PNRR M4.C2.1.1); Japan: Japan Society for the Promotion of Science (JSPS KAKENHI JP22H01227, JSPS KAKENHI JP22H04944, JSPS KAKENHI JP22KK0227, JSPS KAKENHI JP23KK0245); Netherlands: Netherlands Organisation for Scientific Research (NWO Veni 2020 - VI.Veni.202.179); Norway: Research Council of Norway (RCN-314472); Poland: Ministry of Science and Higher Education (IDUB AGH, POB8, D4 no 9722), Polish National Agency for Academic Exchange (PPN/PPO/2020/1/00002/U/00001), Polish National Science Centre (NCN 2021/42/E/ST2/00350, NCN OPUS nr 2022/47/B/ST2/03059, NCN UMO-2019/34/E/ST2/00393, NCN \& H2020 MSCA 945339, UMO-2020/37/B/ST2/01043, UMO-2021/40/C/ST2/00187, UMO-2022/47/O/ST2/00148, UMO-2023/49/B/ST2/04085); Slovenia: Slovenian Research Agency (ARIS grant J1-3010); Spain: Generalitat Valenciana (Artemisa, FEDER, IDIFEDER/2018/048), Ministry of Science and Innovation (MCIN \& NextGenEU PCI2022-135018-2, MICIN \& FEDER PID2021-125273NB, RYC2019-028510-I, RYC2020-030254-I, RYC2021-031273-I, RYC2022-038164-I), PROMETEO and GenT Programmes Generalitat Valenciana (CIDEGENT/2019/027); Sweden: Carl Trygger Foundation (Carl Trygger Foundation CTS 22:2312), Swedish Research Council (Swedish Research Council 2023-04654, VR 2018-00482, VR 2022-03845, VR 2022-04683, VR 2023-03403, VR grant 2021-03651), Knut and Alice Wallenberg Foundation (KAW 2018.0157, KAW 2018.0458, KAW 2019.0447, KAW 2022.0358); Switzerland: Swiss National Science Foundation (SNSF - PCEFP2\_194658); United Kingdom: Leverhulme Trust (Leverhulme Trust RPG-2020-004), Royal Society (NIF-R1-231091); United States of America: U.S. Department of Energy (ECA DE-AC02-76SF00515), Neubauer Family Foundation.

%
%


%
%
%
%
%
%
%

%
\clearpage
\printbibliography
\clearpage
\input{atlas_authlist}
%

%
%
\newpage

\end{document}

%% file: atlas_authlist.tex
 
\begin{flushleft}
\hypersetup{urlcolor=black}
{\Large The ATLAS Collaboration}

\bigskip

\AtlasOrcid[0000-0002-6665-4934]{G.~Aad}$^\textrm{\scriptsize 104}$,
\AtlasOrcid[0000-0001-7616-1554]{E.~Aakvaag}$^\textrm{\scriptsize 17}$,
\AtlasOrcid[0000-0002-5888-2734]{B.~Abbott}$^\textrm{\scriptsize 123}$,
\AtlasOrcid[0000-0002-0287-5869]{S.~Abdelhameed}$^\textrm{\scriptsize 119a}$,
\AtlasOrcid[0000-0002-1002-1652]{K.~Abeling}$^\textrm{\scriptsize 56}$,
\AtlasOrcid[0000-0001-5763-2760]{N.J.~Abicht}$^\textrm{\scriptsize 50}$,
\AtlasOrcid[0000-0002-8496-9294]{S.H.~Abidi}$^\textrm{\scriptsize 30}$,
\AtlasOrcid[0009-0003-6578-220X]{M.~Aboelela}$^\textrm{\scriptsize 45}$,
\AtlasOrcid[0000-0002-9987-2292]{A.~Aboulhorma}$^\textrm{\scriptsize 36e}$,
\AtlasOrcid[0000-0001-5329-6640]{H.~Abramowicz}$^\textrm{\scriptsize 155}$
% ...
\bigskip
\\

% ...
$^{17}$Department for Physics and Technology, University of Bergen, Bergen; Norway.\\
% ...
$^{30}$Physics Department, Brookhaven National Laboratory, Upton NY; United States of America.\\
% ...
$^{36}$$^{(a)}$Facult\'e des Sciences Ain Chock, Universit\'e Hassan II de Casablanca;$^{(b)}$Facult\'{e} des Sciences, Universit\'{e} Ibn-Tofail, K\'{e}nitra;$^{(c)}$Facult\'e des Sciences Semlalia, Universit\'e Cadi Ayyad, LPHEA-Marrakech;$^{(d)}$LPMR, Facult\'e des Sciences, Universit\'e Mohamed Premier, Oujda;$^{(e)}$Facult\'e des sciences, Universit\'e Mohammed V, Rabat;$^{(f)}$Institute of Applied Physics, Mohammed VI Polytechnic University, Ben Guerir; Morocco.\\
% ...
$^{45}$Physics Department, Southern Methodist University, Dallas TX; United States of America.\\
% ...
$^{50}$Fakult\"{a}t Physik , Technische Universit{\"a}t Dortmund, Dortmund; Germany.\\
% ...
$^{56}$II. Physikalisches Institut, Georg-August-Universit\"{a}t G\"ottingen, G\"ottingen; Germany.\\
% ...
$^{104}$CPPM, Aix-Marseille Universit\'e, CNRS/IN2P3, Marseille; France.\\
% ...
$^{119}$$^{(a)}$New York University Abu Dhabi, Abu Dhabi;$^{(b)}$United Arab Emirates University, Al Ain; United Arab Emirates.\\
% ...
$^{123}$Homer L. Dodge Department of Physics and Astronomy, University of Oklahoma, Norman OK; United States of America.\\
% ...
$^{155}$Raymond and Beverly Sackler School of Physics and Astronomy, Tel Aviv University, Tel Aviv; Israel.\\
% ...
$^{a}$ Also Affiliated with an institute covered by a cooperation agreement with CERN.\\
$^{b}$ Also at An-Najah National University, Nablus; Palestine.\\
$^{c}$ Also at Borough of Manhattan Community College, City University of New York, New York NY; United States of America.\\
$^{d}$ Also at Center for Interdisciplinary Research and Innovation (CIRI-AUTH), Thessaloniki; Greece.\\
$^{e}$ Also at Centro Studi e Ricerche Enrico Fermi; Italy.\\
$^{f}$ Also at CERN, Geneva; Switzerland.\\
% ...

\end{flushleft}
